\documentclass{llncs}

\usepackage{mynatbib}
\usepackage{todonotes}
\usepackage{myfloat}
\usepackage{mylistings}
\usepackage{booktabs}
\usepackage{fancyhdr}
\usepackage{rotating}
\usepackage{graphicx}
\usepackage[loose]{subfigure}
\usepackage[utf8x]{inputenc}

\usepackage{mycommands}

\begin{document}

\title{NELL2RDF: Reading the Web,\\ and Publishing it as Linked Data}
\titlerunning{Nell2RDF}  
%
\author{Jos\'e M. Gim\'enez-Garc\'ia\inst{1} \and Ma\'isa Duarte\inst{1} \and Antoine Zimmermann\inst{2} \\ Christophe Gravier\inst{1} \and Estevam R. Hruschka Jr.\inst{3,4} \and Pierre Maret\inst{1}}
\authorrunning{Jos\'e M. Gim\'enez-Garc\'ia et al.} 
%
\tocauthor{Jos\'e M. Gim\'enez-Garc\'ia, Ma\'isa Duarte, and Antoine Zimmermann}
\institute{Univ Lyon, UJM-Saint-\'Etienne, CNRS, Laboratoire Hubert Curien\\ UMR 5516, F-42023 Saint \'Etienne, France\\
\email{\{jose.gimenez.garcia, maisa.duarte, christophe.gravier, pierre.maret\}@univ-st-etienne.fr}
\and
Univ Lyon, MINES Saint-\'Etienne, CNRS, Laboratoire Hubert Curien\\ UMR 5516, F-42023 Saint-\'Etienne, France\\
\email{antoine.zimmermann@emse.fr}
\and
Federal University of Sao Carlos - UFSCar, S\~ao Carlos, Brazil\\
\and
Carnegie Mellon University - CMU, Pittsburgh, United States\\
\email{estevam@cs.cmu.edu}}

\maketitle              

\thispagestyle{fancy}

\begin{abstract}
NELL is a system that continuously reads the Web to extract knowledge in form of entities and relations between them. It has been running since January 2010 and extracted over 50,000,000 candidate statements. NELL's generated data comprises all the candidate statements together with detailed information about how it was generated. This information includes how each component of the system contributed to the extraction of the statement, as well as when that happened and how confident the system is in the veracity of the statement. However, the data is only available in an ad hoc CSV format that makes it difficult to exploit out of the context of NELL. In order to make it more usable for other communities, we adopt Linked Data principles to publish a more standardized, self-describing dataset with rich provenance metadata.
\end{abstract}
\section{Introduction}\label{sec:introduction}

Never-Ending Language Learning (NELL)~\cite{carlson_coupling_2009,mitchell_never-ending_2015} is an autonomous computational system that aims at continually and incrementally learning. NELL has been running for about 7 years in Carnegie Mellon University (US). Currently, NELL has collected over 50 million of candidate beliefs, from with about 3.6 million have been promoted as trustworthy statements. NELL learns from the web and uses an ontology previously created to guide the learning. 
One of the most significant resource contributions of NELL, in addition to the millions of beliefs learned from the Web, is NELL's internal representation (or metadata) for categories, relations and concepts. Such internal representation grows in every iteration, and is used by NELL as a set of different (and constantly updated) \emph{feature vectors} to continuously retrain NELL's learning components and build its own way to understand what is read from the Web. 
\citet{zimmermann_nell2rdf:_2013} published in 2013 a solution to convert NELL's beliefs and ontology into RDF and OWL. However, NELL's internal metadata is not modeled in their work. Thus, the main contribution of this work is to extended the approach to include all the provenance metadata (NELL's internal representation) for each belief. We publish this data using five different representation models: RDF reification~\cite[Sec.~5.3]{brickley_rdf_2014}, N-Ary relations~\cite{noy_defining_2006}, Named Graphs~\cite{carroll_named_2005}, Singleton Properties~\cite{nguyen_dont_2014}, and NdFluents~\cite{gimenez-garcia_ndfluents:_2017}. In addition, we publish not only the promoted beliefs, but also the candidates. As far as we know, this dataset contains more metadata about the statements than any other available dataset in the linked data cloud. This in itself can also be interesting for researchers that seek to manage and exploit meta-knowledge. 

Our intention is to keep this information updated and integrate it on NELL's web page\footnote{\url{http://rtw.ml.cmu.edu/}}.

The rest of the paper is organized as follows: Section \ref{sec:nell} presents NELL and the components it comprises; in Section \ref{sec:nell2rdf} describes the transformation of NELL data and metadata to RDF; Section \ref{sec:dataset} presents the dataset generated in this paper and how it is published; finally, Section \ref{sec:conclusion} provides final remarks and future work.

\section{The Never-Ending Language Learning System}\label{sec:nell}


NELL~\cite{carlson_coupling_2009,mitchell_never-ending_2015} was built based on a new Machine Learning (ML) paradigm, the Never-Ending Learning (NEL). NEL paradigm is a semi-supervised learning~\cite{blum_combining_1998} approach focused on giving the ability to a machine learning system to autonomously use what it has previously learned to continuously become a better learner. NELL is based on a number of coupled components working in parallel. These components read the web and use different approaches to, not only infer new knowledge in the form of beliefs, but also to infer new ways of internally representing the learned beliefs and their properties. Beliefs are divided into candidates and promoted beliefs. In order to be promoted a belief needs to have a confidence score of at least 0.9. 

\begin{enumerate}
	\item \textbf{AliasMatcher} finds relations between entities and their Wikipedia URL on Freebase. It was run only once and is currently not active.
    \item \textbf{CML} \textit{(Coupled Morphologic Learner)} \cite{carlson_toward_2010} is responsible for identifying morphological regularities (such as that words finished in \texttt{burg} could be cities). It makes use of orthographic features of noun phrases (\eg length and number of words, capitalization, prefixes and suffixes).\textit{CMC} is the previous version of this component. 
    \item \textbf{CPL} \textit{(Coupled Pattern Learner)} \cite{carlson_toward_2010} is the component that learns Named Entities (NE) and Textual Patterns (TP) from text in the web pages. Internally, a different implementation was used between 2010 and 2013 that could learn categories and relations together. After that, CPL was splitted in CPL1 and CPL2, the former learning categories and the latter relations, but the distinction is not made in the knowledge base. All the knowledge from CPL1 is promoted promoted only if CPL2 agrees. \ie CPL will extract TPs for categories (\texttt{\_ is a city}, \texttt{city such as \_}, \etc) and for relations (\texttt{arg1 is a city located in arg2}, \texttt{arg1 is the capital of arg2}, \etc). Then, using those TPs, CPL will extract NEs for categories (e.g. \texttt{city(Paris)}, \texttt{city(Annecy)}, \etc) and NE pairs for relations (\texttt{locatedIn(Paris, France)}, \texttt{locatedIn(Annecy, France)}, \etc).
    \item \textbf{KbManipulation} is used to correct some old bugs from NELL's internal indexing knowledge. Several of these bugs should be removed automatically, but NELL has not one automated process for this task yet.
    \item \textbf{LatLong} matches the literal string of Named Entities against a fixed geolocation database. 
    \item  \textbf{LE (Learned Embeddings)} \cite{yang_joint_2016} predicts new categories or relations of entities based on Event and Named Entity extraction 
It creates a feature space where each dimension is a single NELL predicate, and NELL's learned NE (or NE pairs for relations) is used as training examples. LE's process predicts category or relation for NE (or NE pairs) that were not related in the training set.
	\item \textbf{MBL}, also known as \textit{ErrorBasedIntegrator} and \textit{Knowledge Integrator}, is the component responsible for taking the decision of promotion based on the contributions of the other components. \textit{EntityResolverCleanup} is the name used for the same MBL process applied during a big alteration in NELL's knowledge base. In 2010 a big change was made in the NELL’s KB structure to make possible for two words to have different meanings (e.g apple the fruit and Apple the company) and, conversely, for a concept to use different words (e.g Google and Google Inc.).
	\item \textbf{OE} \textit{(Open Eval)} \cite{samadi_openeval:_2013} queries the web and extract small text using predicate instances. OE calculates the score based on the text distance between the instances in a relation.
    \item \textbf{OntologyModifier} is used for any ontology alteration. This component appears in the Knowledge base when a new seed or and ontology extension is manually introduced. 
    \item \textbf{PRA} \textit{(Path Ranking Algorithm)} \cite{gardner_incorporating_2014} is based on Random Walk Inference. PRA analyzes the connections between two categories instances which are the arguments for a relation. This component replaced the old \textit{Ruler Learner} component.
    \item \textbf{RL} \textit{(Rule Learner)} \cite{lao_random_2011} extracts new knowledge using Horn Clauses based on the ontology. Its implementation was based on FOIL \cite{quinlan_foil:_1993}. It can be found in NELL’s KB, but its execution stopped when NELL started to deal with polysemy resolution.
	\item \textbf{SEAL} \textit{(Coupled Set Expander for Any Language)} \cite{wang_language-independent_2007} is the component responsible for extracting knowledge from HTML patterns. It works in a similar way to CPL, but using HTML patterns instead of textual patterns. In the past it was called \textit{CSEAL}, but after some improvements in its performance it changed the name for SEAL.
	\item \textbf{Semparse} \cite{krishnamurthy_joint_2014} combines syntactic parsing from CCGbank (a conversion of the corpus of trees Penn Treebank \cite{marcus_penn_1994}) and distant supervision.
	\item \textbf{SpreadsheetEdits} provides modifications in the NELL's Knowledge base using human feedback. 
\end{enumerate}

Each of of these components, with the exception of \texttt{LE}, output provenance information regarding theirs execution. In the next sections we present how this metadata is modeled in RDF.

\section{Converting NELL to RDF}\label{sec:nell2rdf}

In this section we describe how NELL data  and metadata are transformed into RDF. The first subsection presents how NELL's ontology and beliefs are converted, following the work by \citet{zimmermann_nell2rdf:_2013}; the second subsection describes how we convert the provenance metadata associated with each belief. NELL's Knowledge bases used in this paper for the promoted and candidates beliefs are respectively corresponding to the iterations 1075\footnote{\url{http://rtw.ml.cmu.edu/resources/results/08m/NELL.08m.1075.esv.csv.gz}} and 1070\footnote{\url{http://rtw.ml.cmu.edu/resources/results/08m/NELL.08m.1070.cesv.csv.gz}}. The code is publicly available in GitHub\footnote{\url{https://github.com/WDAqua/nell2rdf}}.

\subsection{Converting NELL's beliefs to RDF}\label{subsec:nell2rdf.data}

NELL's ontology is published as a file with three tab separated values per line, where each line expresses a relationship between categories and other categories, relations, or values used by NELL processes. In order to convert NELL's ontology to RDF each line is transformed into a triple as per \citet{zimmermann_nell2rdf:_2013}. In short, the first and the third values are a pair of categories or relations, or either a category or relation in the first field and a value in the third. The second field is a predicate that indicates the relationship between the two elements.  The transformations can be seen in Table~\ref{tab:ontology}.

\begin{table}
\centering
\caption{NELL’s ontology predicates and their translation in RDFS / OWL (from \cite{zimmermann_nell2rdf:_2013})}
\label{tab:ontology}
\begin{tabular}{l|l}
	\textbf{NELL predicate} & Translation to RDFS / OWL	\\
    \hline
    antireflexive		& rdf:type owl:IrreflexiveProperty \\
    antisymmetric		& antisymmetric Literal(?object,xsd:boolean) \\
    description			& rdfs:comment Literal(?object,@en) \\
    domain				& rdfs:domain Class(?object) \\
    domainwithinrange	& domainWithinRange Literal(?object,xsd:boolean) \\
    generalizations		& rdfs:subClassOf Class(?object) \\
    humanformat 		& humanFormat Literal(?object,xsd:string) \\
    instancetype		& instanceType IRI(?object) \\
    inverse 			& owl:inverseOf ?object \\
    memberofsets 		& \textit{if} ?object \textit{is} rtwcategory \textit{then} rdf:type rdfs:Class \\
    					& \textit{else} ?object \textit{is} rtwrelation \textit{then} rdf:type rdf:Property \\
    mutexpredicates 	& \textit{if} ?subject \textit{is  a}  class  \textit{then} owl:disjointWith ?object \\
    					& \textit{else} ?subject \textit{is a} property \textit{then} owl:propertyDisjointWith ?object \\
    nrofvalues 			& \textit{if} ?object \textit{is} 1 \textit{then} rdf:type owl:FunctionalProperty \\
    populate 			& populate Literal(?object,xsd:boolean) \\
    range 				& rdfs:range ?object \\
    rangewithindomain 	& rangeWithinDomain Literal(?object,xsd:boolean) \\
    visible 			& visible Literal(?object,xsd:boolean) \\
\end{tabular}
\end{table}

NELL's beliefs are also published in tab-separated format, where each line contains a number of fields to express the belief and the associated metadata, such as iteration of promotion, confidence score, or the activity of the components that inferred the belief. All the fields except 4, 5, 6, and 13 are used to convert the beliefs into RDF statements. Table \ref{tab:fields} shows the meaning of each field.  Fields 1, 2, and 3 are converted into the subject, predicate, and object of an RDF statement; the content of fields 7 and 8 create new statements using \texttt{rdf:label} properties; fields 9 and 10 create new triples with the property \texttt{skos:prefLabel}; finally, fields 11 and 12 are used to create triples indicating the types of the subject and the object.
For a more detailed description of this step, refer to \citet{zimmermann_nell2rdf:_2013}.

\begin{table}
\centering
\caption{Description of NELL’s beliefs fields}
\label{tab:fields}
\begin{tabular}{r|l|l}
	\textbf{\#}	&	\textbf{Field}		& \textbf{Description} \\
	\hline
	1	&	Entity						&	Subject of the belief	\\
    2	&	Relation					&	Predicate of the belief	\\
    3	&	Value						&	Object of the belief		\\
    4	&	Iteration					&	Iteration when the belief was promoted, or a list of iterations \\
    	&								&	when the components generated the belief	\\
    5	&	Probability					&	Confidence score of the belief	\\
    6	&	Source						&	MBL activity to promote the belief \\
    7	&	Entity literalStrings		&	Labels of the subject	\\
    8	&	Value literalStrings		&	Labels of the object	\\
    9	&	Best Entity literalString	&	Preferred label of the subject	\\
    10	&	Best Value literalString	&	Preferred label of the object	\\
    11	&	Categories for Entity		&	Classes of the subject	\\
    12	&	Categories for Value		&	Classes of the object	\\
    13	&	Candidate Source			&	Activity of the components that generated the belief\\
\end{tabular}
\end{table}

\subsection{Converting NELL metadata to RDF}\label{subsec:nell2rdf.metadata}

Fields 4, 5, 6, and 13 of each NELL's belief are used to extract the metadata. Each belief is represented by a resource, to which we attach the provenance information. In the promoted beliefs process, field 4 is used to extract the iteration when the belief was promoted, while field 5 gives a confidence score about it. On the other hand, in the candidate beliefs process, fields 4 and 5 contains the iterations when each component generated information about the belief, and the confidence score provided by each of them. Field 6 contains a summary information about the activity of MBL when processing the promoted belief. The complete information from field 6 is a summary of field 13. For that reason, we only process field 13. Finally, in field 13 every activity that took part in generating the statement is parsed.

\begin{sidewaysfigure}
	\centering
	\includegraphics[width=1\textwidth]{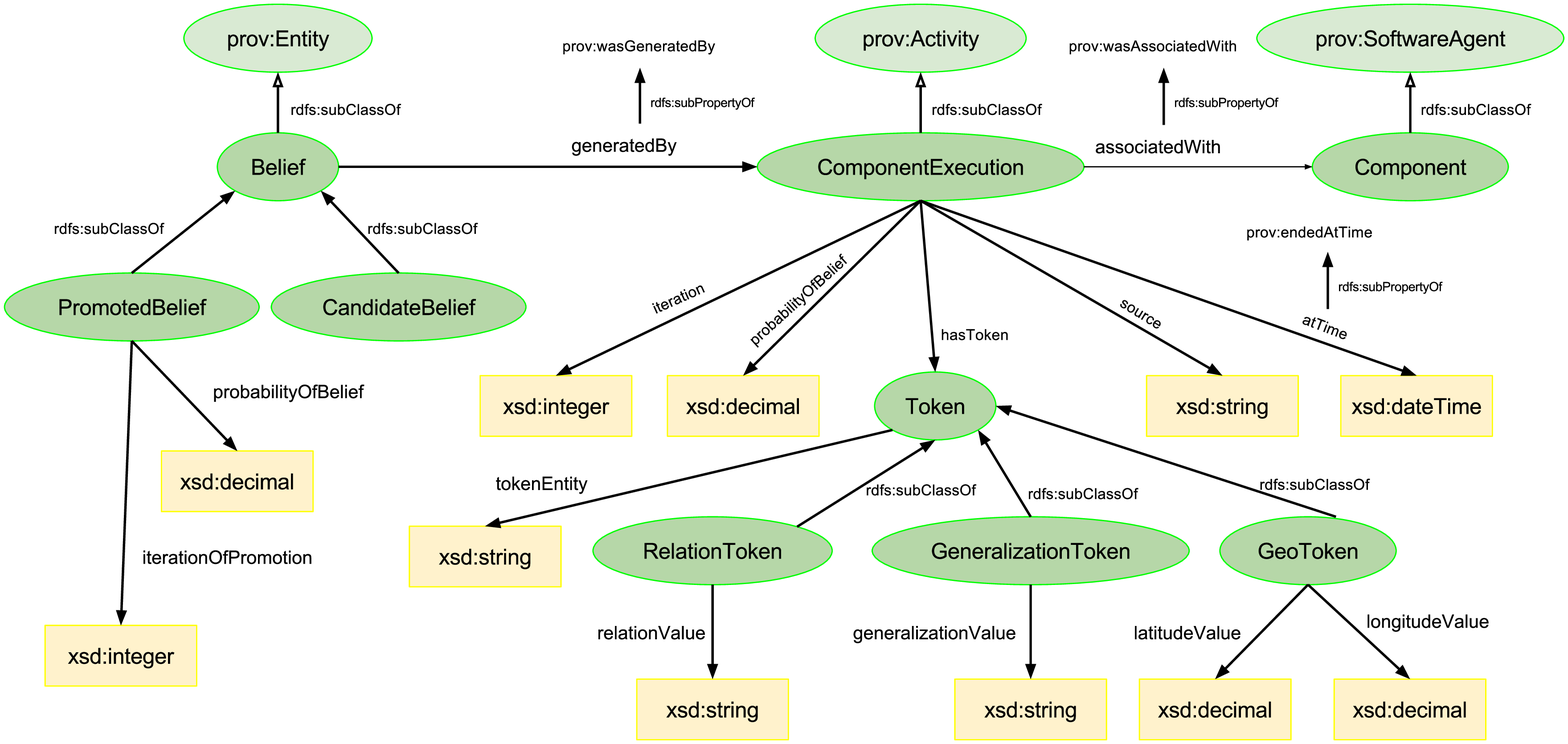}
	\caption{NELL2RDF metadata ontology}
	\label{fig:metadata_ontology}
\end{sidewaysfigure}

The ontology can be seen in Figure \ref{fig:metadata_ontology}. We make use of the PROV-O ontology \cite{lebo_prov-o:_2013} to describe the provenance. Each \texttt{Belief} can be related with one or more \texttt{ComponentExecution} that, in turn, are performed by a \texttt{Component}. If the belief is a \texttt{PromotedBelief}, it has attached its \texttt{iterationOfPromotion} and \texttt{probabilityOfBelief}. The \texttt{ComponentIteration} is related to information about the process: the \texttt{iteration}, \texttt{probabilityOfBelief}, \texttt{Token}, \texttt{source} and \texttt{atTime} (the date and time it was processed). The \texttt{Token} expresses the concepts that the \texttt{Component} is relating. Those concepts can be a pair of entities for a \texttt{RelationToken}, and entity and a class for a \texttt{GeneralizationToken} (note that \texttt{LatLong} component has a different token \texttt{GeoToken}, further described later). Finaly, each component have a \texttt{source} string describing their process for the belief. This string is then further analyzed and translated into a different set if IRIs for each type of component in the subsections below.

The classes of the ontology are described in Table \ref{tab:classes} and properties of the ontology are described in Table \ref{tab:properties}. The classes and properties of each component are described down below.

\begin{table}
\centering
\caption{Description of NELL metadata classes}
\label{tab:classes}
\begin{tabular}{l|l|l}
	\textbf{Class} &	\textbf{rdfs:subClassOf}	& \textbf{Description}	\\
    \hline
	\texttt{Belief}		&	\texttt{prov:Entity}	&	A belief	\\
    \texttt{PromotedBelief}	&	\texttt{Belief}	&	A promoted belief	\\
    \texttt{CandidateBelief}	&	\texttt{Belief}	&	A candidate belief	\\
    \texttt{ComponentExecution}	&	\texttt{prov:Activity}	&	The activity of a component in an iteration \\
    \texttt{Component}	&	\texttt{prov:SoftwareAgent}	&	A component \\
    \texttt{Token}	&	\texttt{owl:Thing}	&	The tuple that was inferred by the activity	\\
    \texttt{RelationToken}	&	\texttt{Token}	&	The tuple \textless Entity,Entity\textgreater~that was 	\\
    	&	&	inferred for a relation \\
    \texttt{GeneralizationToken}	&	\texttt{Token}	&	The tuple \textless Entity,Category\textgreater~that was 	\\
    	&	&	inferred for a generalization \\
        \texttt{GeoToken}	&	\texttt{Token}	&	The tuple \textless Entity,Longitude,Latitude\textgreater  	\\
    &	&	that was inferred for a geografical belief \\
\end{tabular}
\end{table}

\begin{table}
\centering
\caption{Description of NELL metadata properties}
\label{tab:properties}
\begin{tabular}{l|lll}
    \textbf{Property}	& \textbf{rdfs:subPropertyOf}	& \textbf{rdfs:domain}	&	\textbf{rdfs:range}	\\
    	& \multicolumn{3}{|l}{\textbf{Description}} \\
    \hline
    \texttt{generatedBy}	&	\texttt{prov:wasGeneratedBy}	&	\texttt{Belief}	&	\texttt{ComponentIteration}		\\
    & \multicolumn{3}{|l}{The Belief was generated by the iteration of the component} \\
    \texttt{associatedWith}	&	\texttt{prov:wasAssociatedWith}	&	\texttt{ComponentIteration}	&	\texttt{Component}		\\
    & \multicolumn{3}{|l}{The iteration was performed by the component} \\
    \texttt{iterationOfPromotion}	&	\texttt{owl:DatatypeProperty} &	\texttt{PromotedBelief}	& \texttt{xsd:integer}		\\
    & \multicolumn{3}{|l}{iteration in which the component was promoted} \\
    \texttt{probabilityOfBelief}	&	\texttt{owl:DatatypeProperty}	&	\texttt{PromotedBelief}	& \texttt{xsd:decimal}		\\
    & \multicolumn{3}{|l}{Confidence score of the Belief} \\
    \texttt{iteration}	&	\texttt{owl:DatatypeProperty}	&	\texttt{ComponentIteration}	&	\texttt{xsd:integer}	 	\\
    & \multicolumn{3}{|l}{Iteration in which a component performed the activity} \\
    \texttt{probability}	&	\texttt{owl:DatatypeProperty}	&	\texttt{ComponentIteration}	&	\texttt{xsd:decimal}		\\
    & \multicolumn{3}{|l}{Confidence score given by the component} \\
    \texttt{hasToken}	&	owl:ObjectProperty	&	\texttt{ComponentIteration}	& \texttt{Token} \\
    & \multicolumn{3}{|l}{The concepts that the component is relating}		\\
    \texttt{source}	&	\texttt{owl:DatatypeProperty}	&	\texttt{ComponentIteration}	&	\texttt{xsd:string}		\\
    & \multicolumn{3}{|l}{Data that was used by the component in the activity} \\
    \texttt{atTime}	&	\texttt{owl:DatatypeProperty}	&	\texttt{ComponentIteration}	&	\texttt{xsd:dateTime}		\\
    & \multicolumn{3}{|l}{Date and time when the component execution was performed} \\
    \texttt{tokenEntity}		&	\texttt{owl:DatatypeProperty}	&	\texttt{Token}	&	\texttt{xsd:string}		\\
    & \multicolumn{3}{|l}{Entity on which the data was inferred} \\
    \texttt{relationValue}	&	\texttt{owl:DatatypeProperty}	&	\texttt{RelationToken}	&	\texttt{xsd:string}		\\
    & \multicolumn{3}{|l}{Entity related the entity appointed by \texttt{tokenEntity}} \\
    \texttt{generalizationValue}	&	\texttt{owl:DatatypeProperty}	&	\texttt{GeneralizationToken}	&	\texttt{xsd:string}		\\
    & \multicolumn{3}{|l}{Class of the entity appointed by \texttt{tokenEntity}} \\	
\end{tabular}
\end{table}

\begin{figure}
	\centering
    \includegraphics[height=0.30\linewidth]{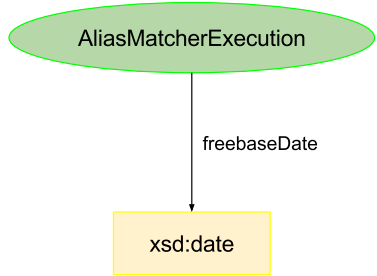}
    \caption{AliasMatcherExecution metadata ontology}
    \label{fig:aliasmatcherexecution}
\end{figure}

\paragraph{AliasMatcher} execution is denoted by a resource of class \texttt{AliasMatcherExecution}, and includes the date when the data was extracted from Freebase using the property \texttt{freebaseDate}. The added ontology can be seen in Figure~\ref{fig:aliasmatcherexecution}.



\begin{figure}
	\centering
    \includegraphics[height=0.50\linewidth]{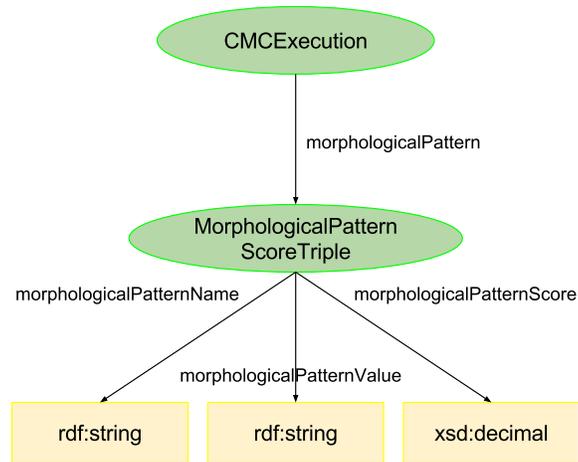}
    \caption{CMC metadata ontology}
    \label{fig:cmcexecution}
\end{figure}

\paragraph{CMC} execution is denoted by a resource of class \texttt{CMCExecution}. A number of morphological patterns \texttt{MorphologicalPatternScoreTriple} are attached to it, each one containing a name, a value, and a confidence score. The properties used can be seen in Table~\ref{tab:properties.cmc}, while the ontology diagram is shown in Figure~\ref{fig:cmcexecution}.


\begin{table}
\centering
\caption{Description of CMC metadata properties}
\label{tab:properties.cmc}
\begin{tabular}{l|ll}
    \textbf{Property}	& \textbf{rdfs:domain}	&	\textbf{rdfs:range}	\\
    	& \multicolumn{2}{|l}{\textbf{Description}} \\
    \hline
    \texttt{morphologicalPattern}	&	\texttt{CMCExecution}	&	\texttt{MorphologicalPatternScoreTriple}		\\
    & \multicolumn{2}{|l}{One of the morphological patterns used by \texttt{CMC}} \\
     \hline
    \texttt{morphologicalPatternName}	&	\texttt{MorphologicalPatternScoreTriple}	&	\texttt{xsd:string}		\\
    & \multicolumn{2}{|l}{Name of the morphological pattern (\ie prefix, suffix, etc.)} \\
     \hline
    \texttt{morphologicalPatternValue}	&	\texttt{MorphologicalPatternScoreTriple}	&	\texttt{xsd:string}		\\
    & \multicolumn{2}{|l}{Value of the morphological pattern (\ie prefix = Saint and suffix = burgh)} \\
     \hline
    \texttt{morphologicalPatternScore}	&	\texttt{MorphologicalPatternScoreTriple}	&	\texttt{xsd:decimal}		\\
    & \multicolumn{2}{|l}{Score of the morphological pattern} \\
\end{tabular}
\end{table}

\begin{figure}
	\centering
    \includegraphics[height=0.50\linewidth]{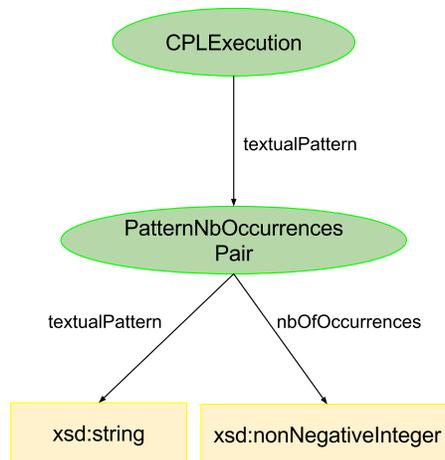}
    \caption{CPL metadata ontology}
    \label{fig:cplexecution}
\end{figure}

\paragraph{CPL} execution is denoted by a resource of class \texttt{CPLExecution}. It contains a series of textual patterns \texttt{patternOccurrences}, each one with a literal that describes the pattern, and the number of times it has occurred in the NELL's data source. The properties used are described in Table~\ref{tab:properties.cpl}, and the diagram for the ontology is shown in Figure~\ref{fig:cplexecution}.


\begin{table}
\centering
\caption{Description of CPL metadata properties}
\label{tab:properties.cpl}
\begin{tabular}{l|ll}
    \textbf{Property}	& \textbf{rdfs:domain}	&	\textbf{rdfs:range}	\\
    	& \multicolumn{2}{|l}{\textbf{Description}} \\
    \hline
    \texttt{patternOccurrences}	& \texttt{CPLExecution}	&	\texttt{PatternNbOfOccurrencesPair}		\\
    & \multicolumn{2}{|l}{One of the textual patterns used by \texttt{CPL}} \\
    \hline
    \texttt{textualPattern}	& \texttt{PatternNbOfOccurrencesPair}	&	\texttt{xsd:string}		\\
    & \multicolumn{2}{|l}{Textual pattern in the form of a sentence} \\
    \hline
    \texttt{nbOfOccurrences}	& \texttt{PatternNbOfOccurrencesPair}	&	\texttt{xsd:nonNegativeInteger}		\\
    & \multicolumn{2}{|l}{Number of times it has occurred in the NELL's source data} \\
\end{tabular}
\end{table}

\begin{figure}
	\centering
    \includegraphics[height=0.30\linewidth]{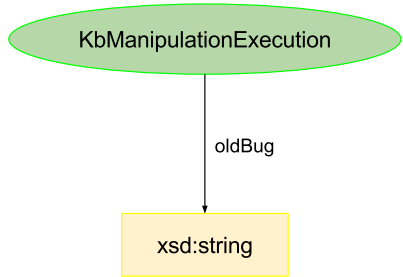}
    \caption{KbManipulation metadata ontology}
    \label{fig:kbmanipulationexecution}
\end{figure}

\paragraph{KbManipulation} execution is denoted by a resource of class \texttt{KbManipulationExecution}. Ir contains the bug \texttt{oldBug} that was manually fixed. Its shown in Figure~\ref{fig:kbmanipulationexecution}.



\begin{figure}
	\centering
    \includegraphics[height=0.50\linewidth]{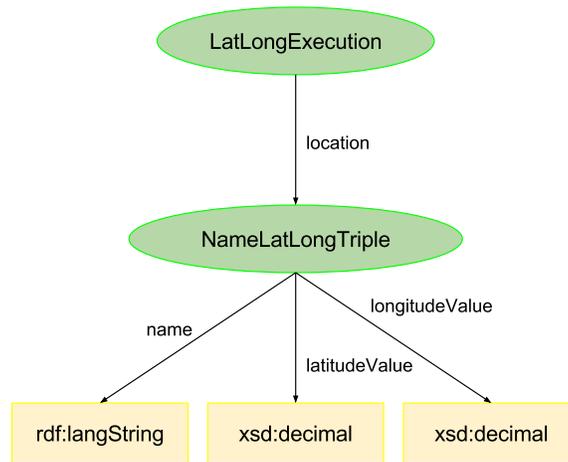}
    \caption{LatLong metadata ontology}
    \label{fig:latlongexecution}
\end{figure}

\paragraph{LatLong} execution is denoted by a resource of class \texttt{LatLongExecution}. It contains a list of locations \texttt{NameLatLongTriple} that were used to infer the belief. Each one containing the \texttt{name} and the latitude and longitude values. This execution has also its own token \texttt{GeoToken} with the latitude and longitude values reusing the same properties. The properties are detailed in Table~\ref{tab:properties.latlong}, and the ontology diagram is shown in Figure~\ref{fig:latlongexecution}.

    

\begin{table}
\centering
\caption{Description of LatLong metadata properties}
\label{tab:properties.latlong}
\begin{tabular}{l|ll}
    \textbf{Property}	& \textbf{rdfs:domain}	&	\textbf{rdfs:range}	\\
    	& \multicolumn{2}{|l}{\textbf{Description}} \\
    \hline
    \texttt{location}	&	\texttt{LatLongExecution}	&	\texttt{NameLatLongTriple}		\\
    & \multicolumn{2}{|l}{One of the locations used by \texttt{Latlong}} \\
    \texttt{name}	&	\texttt{NameLatLongTriple}	&	\texttt{rdf:langString}		\\
    & \multicolumn{2}{|l}{Name of the location} \\
    \texttt{latitudeValue}	&	\texttt{NameLatLongTriple}	&	\texttt{xsd:decimal}		\\
    & \multicolumn{2}{|l}{Latitude of the location} \\
    \texttt{longitudeValue}	&	\texttt{NameLatLongTriple}	&	\texttt{xsd:decimal}		\\
    & \multicolumn{2}{|l}{Longitude of the location} \\
\end{tabular}
\end{table}

\paragraph{LE} execution is denoted by a resource of class \texttt{LEExecution}. It does not contain any additional triples.

\begin{figure}
	\centering
    \includegraphics[height=0.40\linewidth]{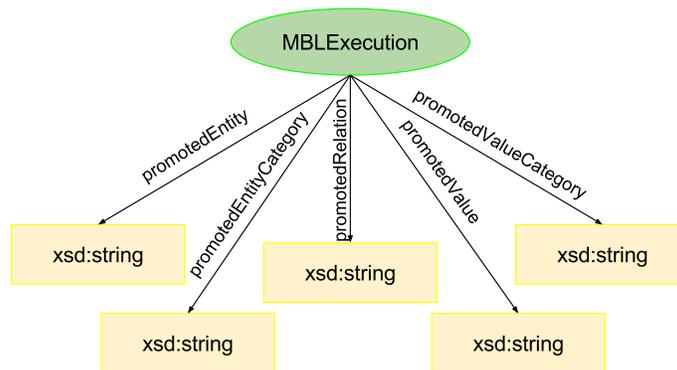}
    \caption{MBL metadata ontology}
    \label{fig:mblexecution}
\end{figure}

\paragraph{MBL} execution is denoted by a resource of class \texttt{MBLExecution}. It contains the entities and the categories of the other belief that was used to promote this one. The properties used are described in Table~\ref{tab:properties.mbl}, and the ontology diagram is shown in Figure~\ref{fig:mblexecution}.


\begin{table}
\centering
\caption{Description of MBL metadata properties}
\label{tab:properties.mbl}
\begin{tabular}{l|ll}
    \textbf{Property}	& \textbf{rdfs:domain}	&	\textbf{rdfs:range}	\\
    	& \multicolumn{2}{|l}{\textbf{Description}} \\
    \hline
    \texttt{promotedEntity}	& \texttt{MBLExecution}	&	\texttt{xsd:string}		\\
    & \multicolumn{2}{|l}{Entity of a belief previously promoted} \\
    \hline
    \texttt{promotedEntityCategory}	& \texttt{MBLExecution}	&	\texttt{xsd:string}		\\
    & \multicolumn{2}{|l}{Category of the entity of the promoted belief} \\
    \hline
    \texttt{promotedRelation}	& \texttt{MBLExecution}	&	\texttt{xsd:string}		\\
    & \multicolumn{2}{|l}{Relation of the promoted belief} \\
    \hline
    \texttt{promotedValue}	& \texttt{MBLExecution}	&	\texttt{xsd:string}		\\
    & \multicolumn{2}{|l}{Value of the promoted belief} \\
    \hline
    \texttt{promotedValueCategory}	& \texttt{MBLExecution}	&	\texttt{xsd:string}		\\
    & \multicolumn{2}{|l}{Category of the promoted belief, if applicable} \\
\end{tabular}
\end{table}

\begin{figure}
	\centering
    \includegraphics[height=0.50\linewidth]{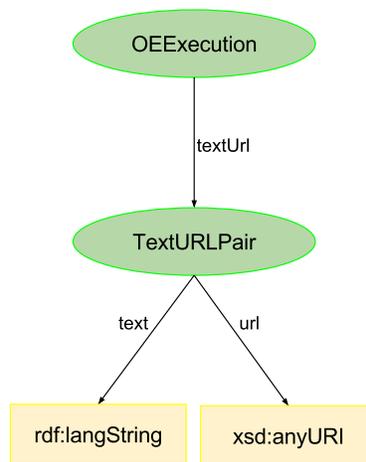}
    \caption{OE metadata ontology}
    \label{fig:oeexecution}
\end{figure}

\paragraph{OE} execution is denoted by a resource of class \texttt{OEExecution}. It contains a set of pairs \texttt{TextUrlPair}, each one including the sentence that was used to infer the belief, and the URL from where it was extracted. The properties used can be found in Table~\ref{tab:properties.oe}, and the ontology diagram in Figure~\ref{fig:oeexecution}.


\begin{table}
\centering
\caption{Description of OE metadata properties}
\label{tab:properties.oe}
\begin{tabular}{l|ll}
    \textbf{Property}	& \textbf{rdfs:domain}	&	\textbf{rdfs:range}	\\
    	& \multicolumn{2}{|l}{\textbf{Description}} \\
    \hline
    \texttt{textUrl}	& \texttt{OEExecution}	&	\texttt{TextUrlPair}		\\
    & \multicolumn{2}{|l}{One of the pairs \textless text, url\textgreater~used by \texttt{OE}} \\
    \texttt{text}	& \texttt{TextUrlPair}	&	\texttt{rdf:langString}		\\
    & \multicolumn{2}{|l}{Text extracted from the web} \\
    \texttt{url}	&	&	\texttt{xsd:anyURI}		\\
    & \multicolumn{2}{|l}{Web page where the text was extracted} \\
\end{tabular}
\end{table}

\begin{figure}
	\centering
    \includegraphics[height=0.30\linewidth]{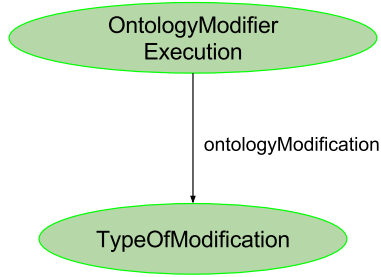}
    \caption{OntologyModifier metadata ontology}
    \label{fig:ontologymodifierexecution}
\end{figure}

\paragraph{OntologyModifier} execution is denoted by a resource of class \texttt{OntologyModifierExecution}. It contains the \texttt{ontologyModification}, which can be either a modification of a category or a modification of a relation. The ontology diagram can be seen in Figure~\ref{fig:ontologymodifierexecution}.

%

\begin{figure}
	\centering
    \includegraphics[height=0.50\linewidth]{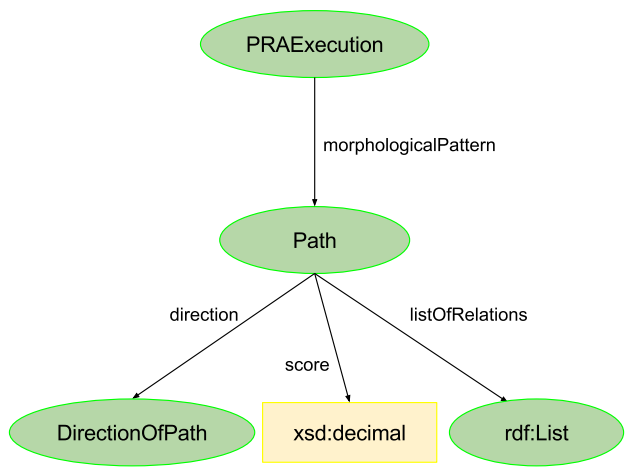}
    \caption{PRA metadata ontology}
    \label{fig:praexecution}
\end{figure}

\paragraph{PRA} execution is denoted by a resource of class \texttt{PRAExecution}. It includes a series of \texttt{Path} resources describing the path followed in NELL dataset to infer the belief. Each \texttt{Path} includes its direction and a confidence score, along with a list of relations followed. The properties used can be seen in Table~\ref{tab:properties.pra}, while the ontology diagram is shown in Figure~\ref{fig:praexecution}.


\begin{table}
\centering
\caption{Description of PRA metadata properties}
\label{tab:properties.pra}
\begin{tabular}{l|ll}
    \textbf{Property}	& \textbf{rdfs:domain}	&	\textbf{rdfs:range}	\\
    	& \multicolumn{2}{|l}{\textbf{Description}} \\
    \hline
    \texttt{relationPath}	& \texttt{PRAExecution}	&	\texttt{Path}		\\
    & \multicolumn{2}{|l}{Relation path that entails the belief} \\
    \texttt{direction}	& \texttt{Path}	&	\texttt{DirectionOfPath}		\\
    & \multicolumn{2}{|l}{Direction of the path} \\
    \texttt{score}	& \texttt{Path}	&	\texttt{xsd:decimal}		\\
        & \multicolumn{2}{|l}{Score assigned to the entailment} \\
    \texttt{listOfRelations}	& \texttt{Path}	&	\texttt{rdf:List}		\\
    & \multicolumn{2}{|l}{Ordered list of relations in the path} \\
\end{tabular}
\end{table}

\begin{figure}
	\centering
    \includegraphics[width=1\linewidth]{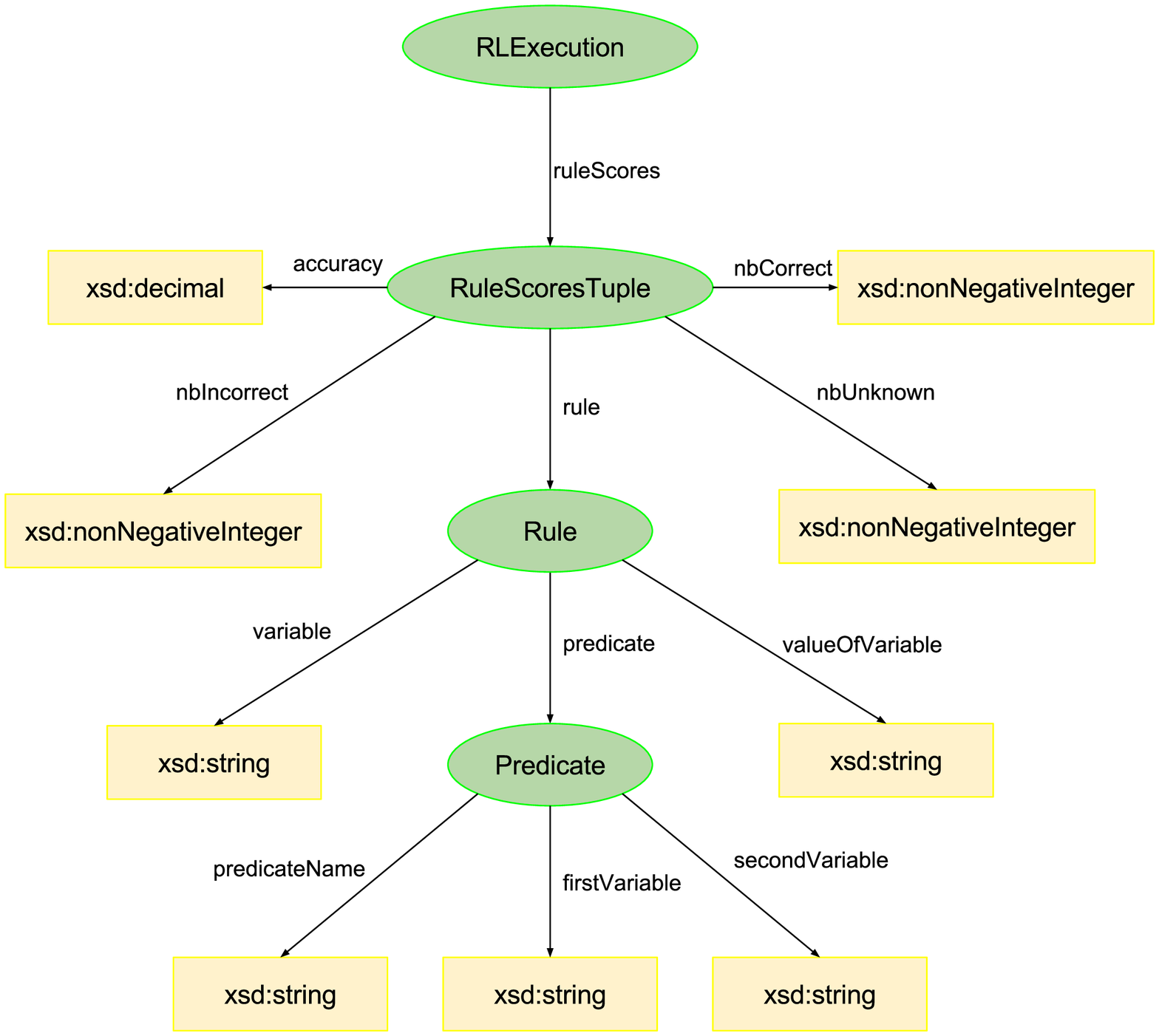}
    \caption{RL metadata ontology}
    \label{fig:rlexecution}
\end{figure}

\paragraph{RL} execution is denoted by a resource of class \texttt{RLExecution}. It contains a resource \texttt{RuleScoresTuple} that contains the \texttt{Rule} and a set of scores indicating the confidence, and the number of beliefs that are estimated to be correctly and incorrectly inferred (and the number of inferred beliefs for which it is not known if they are correct or not) with that rule. The rule itself contains the variables and their values, and the predicates that are part of it. Each \texttt{Predicate} includes the name of the predicate and the two variables it uses. The complete list of properties can be found in table~\ref{tab:properties.rl}. The ontology diagram is presented in Figure~\ref{fig:rlexecution}.


\begin{table}
\centering
\caption{Description of RL metadata properties}
\label{tab:properties.rl}
\begin{tabular}{l|ll}
    \textbf{Property}	& \textbf{rdfs:domain}	&	\textbf{rdfs:range}	\\
    	& \multicolumn{2}{|l}{\textbf{Description}} \\
    \hline
    \texttt{ruleScores}	& \texttt{RLExecution}	&	\texttt{RuleScoresTuple}		\\
    & \multicolumn{2}{|l}{The rule and set of scores used by \texttt{RL}} \\
    \texttt{rule}	& \texttt{RuleScoresTuple}	&	\texttt{Rule}		\\
    & \multicolumn{2}{|l}{The rule \texttt{RL} used to infer the belief, in the form of horn clauses} \\
    \texttt{accuracy}	& \texttt{RuleScoresTuple}	&	\texttt{xsd:decimal}		\\
    & \multicolumn{2}{|l}{Estimated accuracy of the rule in NELL} \\
    \texttt{nbCorrect}	& \texttt{RuleScoresTuple}	&	\texttt{xsd:nonNegativeInteger}		\\
    & \multicolumn{2}{|l}{Estimated number of correct beliefs created by the rule} \\
    \texttt{nbIncorrect}	& \texttt{RuleScoresTuple}	&	\texttt{xsd:nonNegativeInteger}		\\
    & \multicolumn{2}{|l}{Estimated number of incorrect beliefs created by the rule} \\
    \texttt{nbUnknown}	& \texttt{RuleScoresTuple}	&	\texttt{xsd:nonNegativeInteger}		\\
    & \multicolumn{2}{|l}{Number of rules created by the rules with no known correctness} \\
    \texttt{variable}	& \texttt{Rule}	&	\texttt{xsd:string}		\\
    & \multicolumn{2}{|l}{One of the variables that appear in the rule} \\
    \texttt{valueOfVariable}	& \texttt{Rule}	&	\texttt{xsd:string}		\\
    & \multicolumn{2}{|l}{Value of the variable inferred by the rule} \\
    \texttt{predicate}	& \texttt{Rule}	&	\texttt{Predicate}		\\
    & \multicolumn{2}{|l}{One of the predicates that appear in the rule} \\
    \texttt{predicateName}	& \texttt{Predicate}	&	\texttt{xsd:string}		\\
    & \multicolumn{2}{|l}{Name of the predicate} \\
    \texttt{firstVariable}	& \texttt{Predicate}	&	\texttt{xsd:string}		\\
    & \multicolumn{2}{|l}{First variable of the predicate} \\
    \texttt{secondVariable}	& \texttt{Predicate}	&	\texttt{xsd:string}		\\
    & \multicolumn{2}{|l}{Second variable of the predicate} \\
\end{tabular}
\end{table}

\begin{figure}
	\centering
    \includegraphics[height=0.30\linewidth]{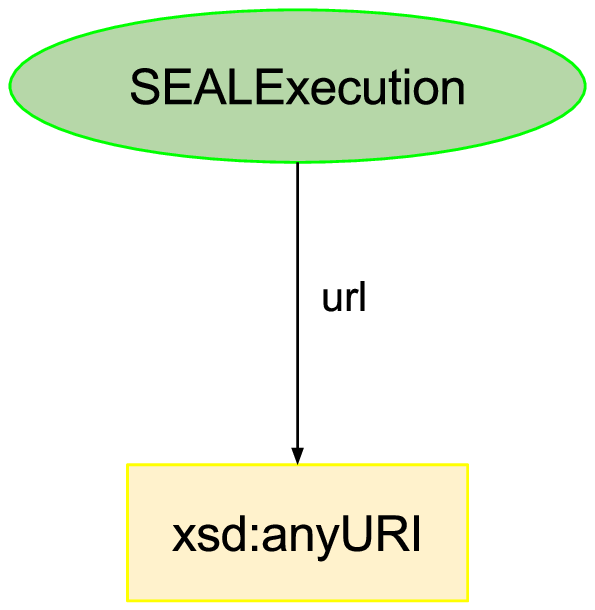}
    \caption{SEAL metadata ontology}
    \label{fig:sealexecution}
\end{figure}

\paragraph{SEAL} execution is denoted by a resource of class \texttt{SEALExecution}. It includes the URL it used with the property \texttt{url}. The ontology diagram can be seen in Figure~\ref{fig:sealexecution}.



\begin{figure}
	\centering
    \includegraphics[height=0.30\linewidth]{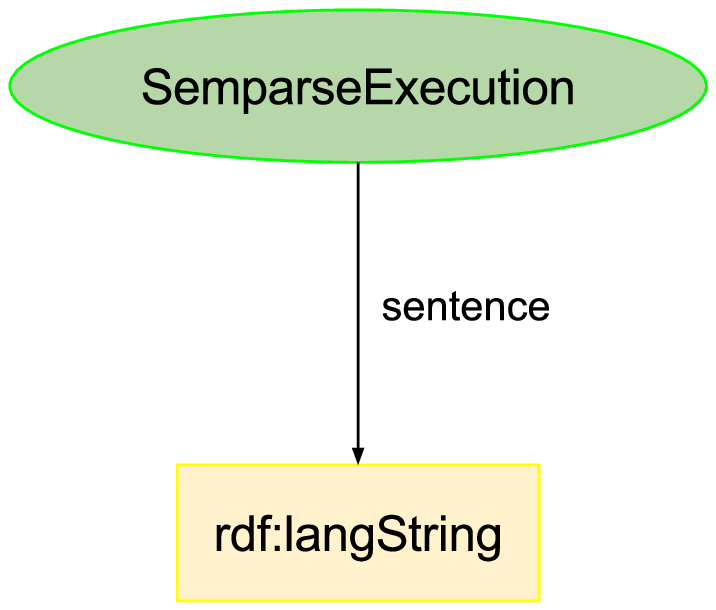}
    \caption{Semparse metadata ontology}
    \label{fig:semparseexecution}
\end{figure}

\paragraph{Semparse} execution is denoted by a resource of class \texttt{SemparseExecution}. It includes a literal with the sentence used during it, using the property \texttt{sentence}. The ontology diagram can be seen in Figure~\ref{fig:semparseexecution}.



\begin{figure}
	\centering
    \includegraphics[height=0.40\linewidth]{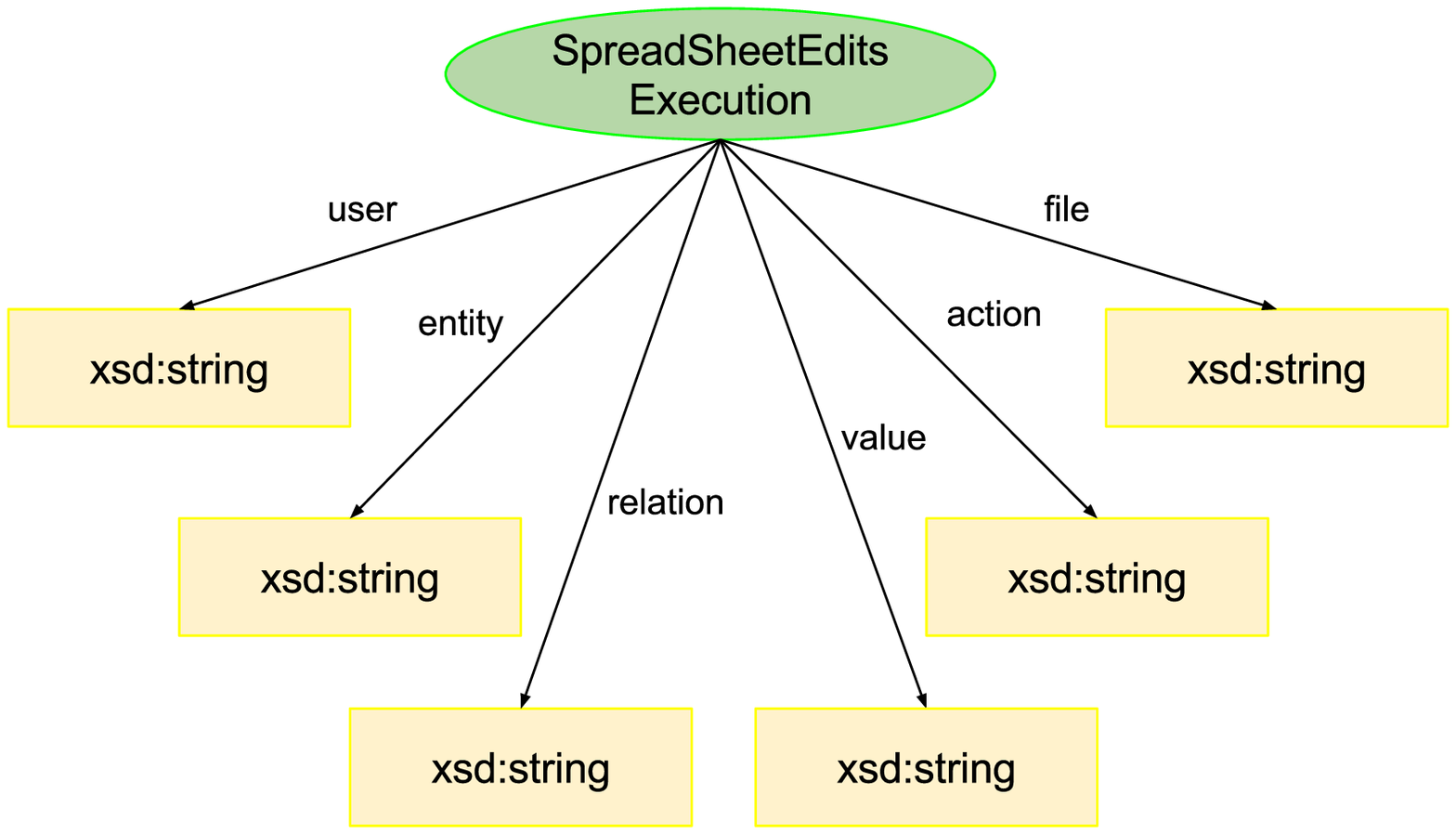}
    \caption{SpreadsheetEdits metadata ontology}
    \label{fig:spreadsheeteditsexecution}
\end{figure}

\paragraph{SpreadsheetEdits} execution is denoted by a resource of class \texttt{SpreadsheetEditsExecution}. It contains a set of literals describing the user who made the modification, the file used as input, the action made, and the modified entity, relation, and value. The list of properties can be seen in Table~\ref{tab:properties.spreadsheetedits}, while the ontology diagram is shown in Figure~\ref{fig:spreadsheeteditsexecution}.


\begin{table}
\centering
\caption{Description of SpreadsheetEdits metadata properties}
\label{tab:properties.spreadsheetedits}
\begin{tabular}{l|ll}
    \textbf{Property}	& \textbf{rdfs:domain}	&	\textbf{rdfs:range}	\\
    	& \multicolumn{2}{|l}{\textbf{Description}} \\
    \hline
    \texttt{user}	& \texttt{SpreadsheetEditsExecution}	&	\texttt{xsd:string}		\\
    & \multicolumn{2}{|l}{User that made the modification} \\
    \texttt{entity}	& \texttt{SpreadsheetEditsExecution}	&	\texttt{xsd:string}		\\
    & \multicolumn{2}{|l}{Entity of the belief affected by the modification} \\
    \texttt{relation}	& \texttt{SpreadsheetEditsExecution}	&	\texttt{xsd:string}		\\
    & \multicolumn{2}{|l}{Relation of the belief affected by the modification} \\
    \texttt{value}	& \texttt{SpreadsheetEditsExecution}	&	\texttt{xsd:string}		\\
    & \multicolumn{2}{|l}{Value of the belief affected by the modification} \\
    \texttt{action}	& \texttt{SpreadsheetEditsExecution}	&	\texttt{xsd:string}		\\
    & \multicolumn{2}{|l}{Action made in the modification} \\
    \texttt{file}	& \texttt{SpreadsheetEditsExecution}	&	\texttt{xsd:string}		\\
    & \multicolumn{2}{|l}{File where the modification was saved and then read by \texttt{SpreadsheetEdits}} \\
\end{tabular}
\end{table}

\section{The NELL2RDF Dataset}\label{sec:dataset}

The current version of NELL2RDF updates the promoted beliefs to the last version, adding the provenance triples about them. It also adds the candidate beliefs and their corresponding provenance triples. We provide the dumps for the promoted beliefs\footnote{\url{https://w3id.org/nellrdf/nellrdf.promoted.n3.gz}} and the candidate beliefs\footnote{\url{https://w3id.org/nellrdf/nellrdf.candidates.n3.gz}}. The ontologies for the beliefs\footnote{\url{https://w3id.org/nellrdf/ontology/nellrdf.ontology.n3}} and the provenance metadata\footnote{\url{https://w3id.org/nellrdf/provenance/ontology/nellrdf.ontology.n3}} is common for both dumps. Metadata about the dataset\footnote{\url{https://w3id.org/nellrdf/metadata/nellrdf.metadata.n3}} is modeled using VoID and DCAT vocabularies.

\begin{sidewaysfigure}
	\centering
	\subfigure[Original Triple]{
		\centering
		\includegraphics[width=0.25\linewidth]{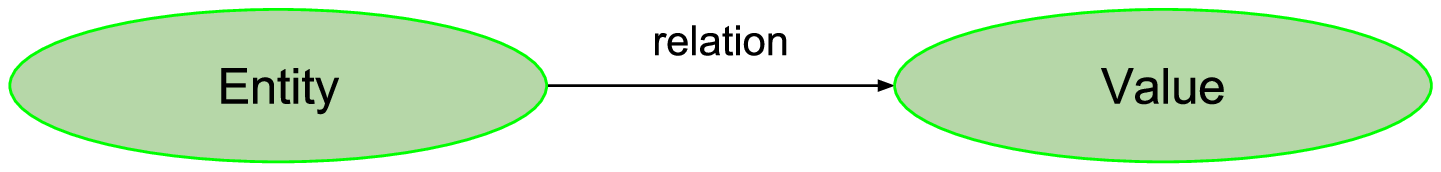}
		\label{subfig:belief}
	}
    \subfigure[RDF Reification]{
		\centering
		\includegraphics[width=0.40\linewidth]{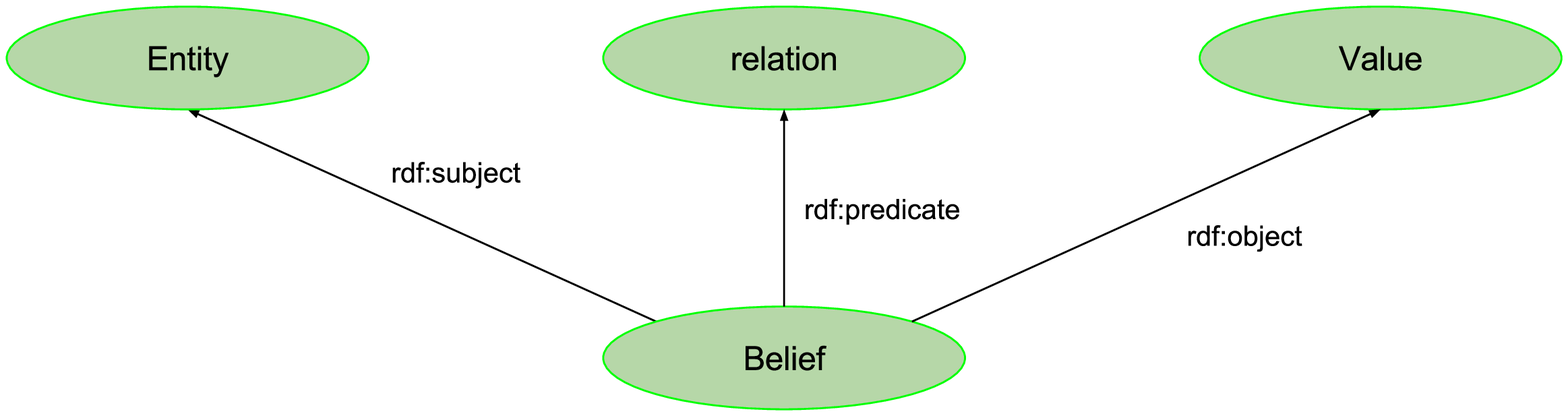}
		\label{subfig:reification}
	}
    \subfigure[Singleton Property]{
		\centering
		\includegraphics[width=0.25\linewidth]{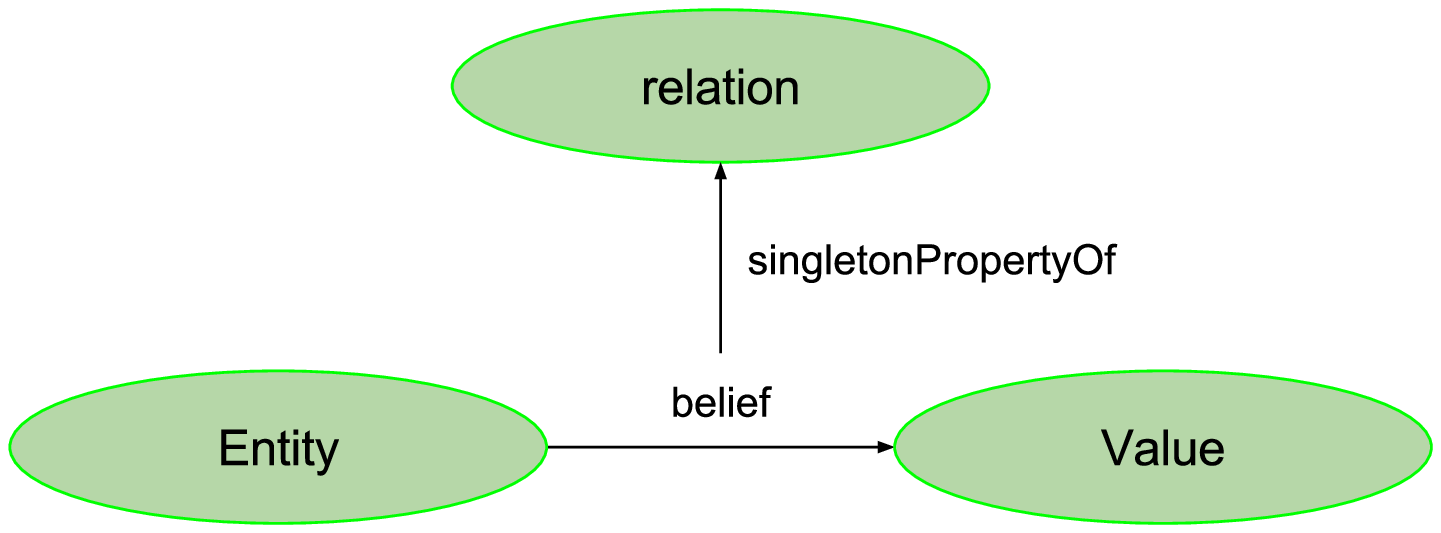}
		\label{subfig:sp}
	}
    \subfigure[Named Graphs]{
		\centering
		\includegraphics[width=0.25\linewidth]{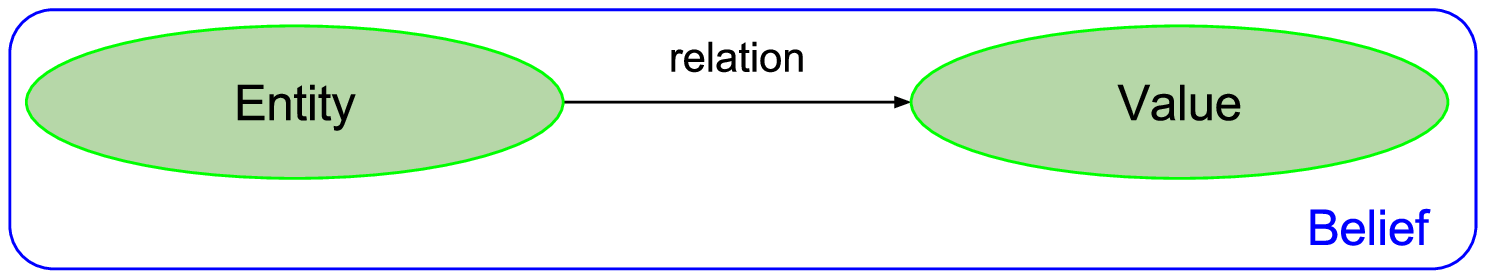}
		\label{subfig:namedgraph}
	}
     \subfigure[N-Ary Properties]{
		\centering
		\includegraphics[width=0.40\linewidth]{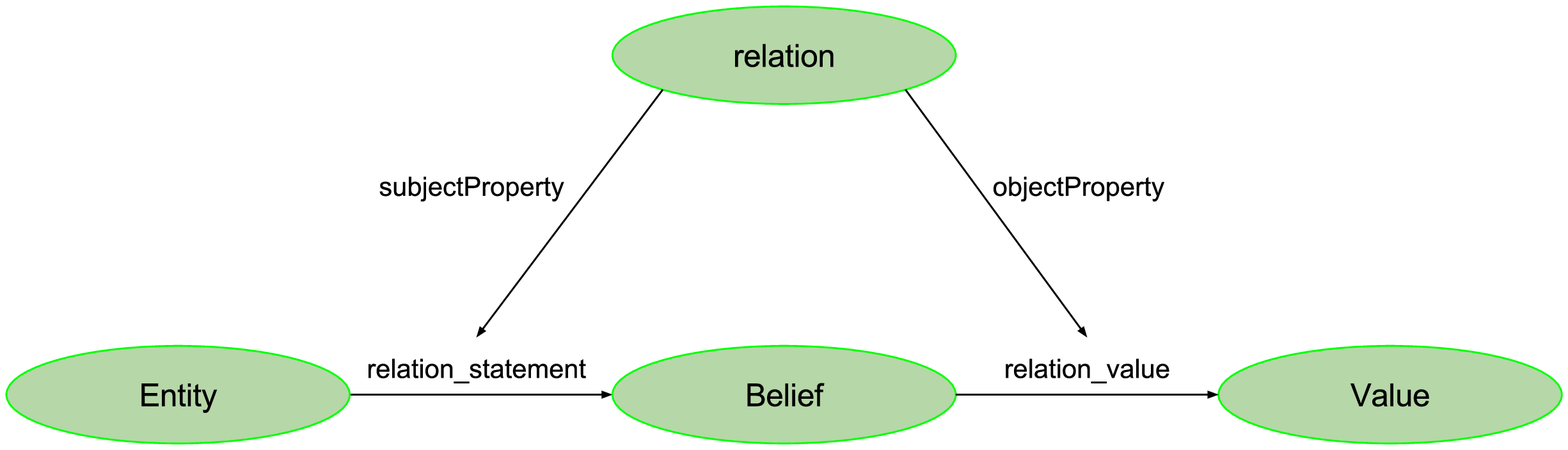}
		\label{subfig:nary}
	}
    \subfigure[NdFluents]{
		\centering
		\includegraphics[width=0.25\linewidth]{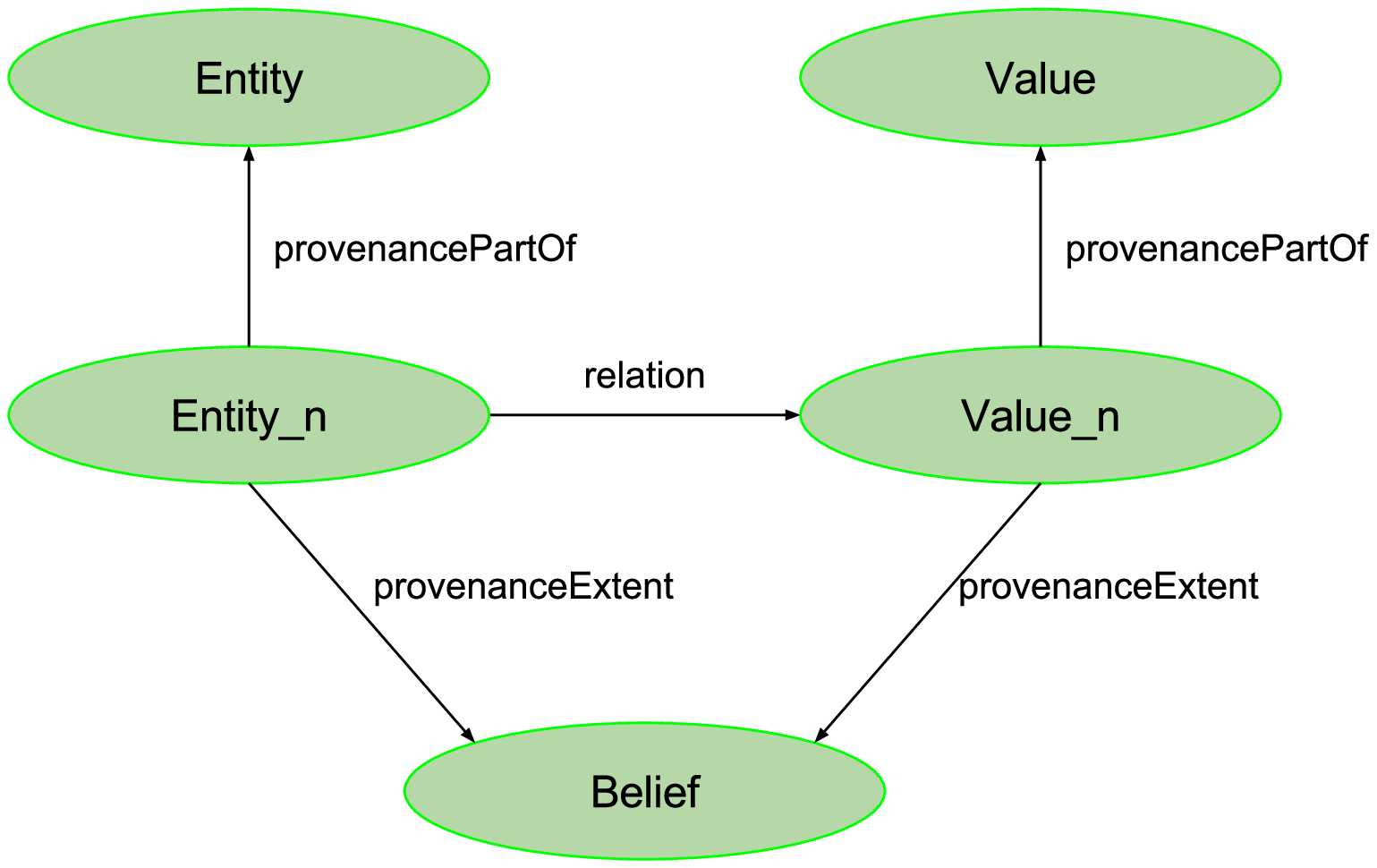}
		\label{subfig:ndfluents}
	}
    \caption{Reification models}
    \label{fig:6figures}
\end{sidewaysfigure}

In order to attach the metadata to each belief, we need to reify the statement into a resource. We follow five different models, described down below. A graphical representation of the models is shown in Figure~\ref{fig:6figures}. A summary of the triples and resources of each model can be seen in Table~\ref{tab:models}.

\begin{itemize}
	\item \emph{RDF Reification}~\cite[Sec.~5.3]{brickley_rdf_2014} represents the statement using a resource, and then creates triples to indicate the subject, predicate and object of the statement.
    \item \emph{N-Ary relations}~\cite{noy_defining_2006}: This model creates a new resource that identifies the relation and connects subject and object using different design patterns. Wikidata\footnote{\url{https://www.wikidata.org}} makes use of this model of annotation.
    \item \emph{Named Graphs}~\cite{carroll_named_2005}: A forth element is added to each triple, that can be used to identify a triple or set of triples later on. This model is used by Nano-publications~\cite{mons_nano-publication_2009}.
    \item \emph{The Singleton Property}~\cite{nguyen_dont_2014} creates a unique property for each triple, related to the original one. It defines its own semantics that extend RDF, RDFS.
    \item \emph{NdFluents}~\cite{gimenez-garcia_ndfluents:_2017} creates a unique version of the subject and the object (in the case it is not a literal) of the triple, and attaches them to the original resources and the context of the statement.
\end{itemize}

\begin{table}
\centering
\caption{Summary of dataset stats for each model}
\label{tab:models}
\begin{tabular}{r|rr|rr|rr|}
	& \multicolumn{2}{c|}{\textbf{Promoted}} & \multicolumn{2}{c|}{\textbf{Candidates}} & \multicolumn{2}{c|}{\textbf{Total}} \\
    \textbf{Model}	& \textbf{Size}	&	\textbf{Triples}	& \textbf{Size}	&	\textbf{Triples}	& \textbf{Size}	&	\textbf{Triples}	\\
    \hline
    \textbf{W/O metadata}		&	2.99GB	&	0.02B	&	162GB	&	1.45B	&	165GB	&	1.48B	\\
    \textbf{RDF Reification}	&	50.9GB	&	0.24B	&	776GB	&	4.50B	&	827GB	&	4.74B	\\
    \textbf{N-Ary Relations}	&	50.7GB	&	0.24B	&	770GB	&	4.50B	&	821GB	&	4.74B	\\
    \textbf{Named Graphs}		&	49.8GB	&	0.24B	&	727GB	&	4.24B	&	777GB	&	4.48B	\\
    \textbf{Singleton Property}	&	49.8GB	&	0.24B	&	xxxGB	&	x.xxB	&	xxxGB	&	x.xxB	\\
    \textbf{NdFluents}			&	51.3GB	&	0.25B	&	xxxGB	&	x.xxB	&	xxxGB	&	x.xxB	\\
\end{tabular}
\end{table}

\section{Discussion and Future Work}\label{sec:conclusion}

In this work we present the conversion of both data and metadata from NELL into RDF. It presents a thesaurus of entities and binary relations between them, as well as a number of lexicalizations for each entity. It also includes detailed provenance metadata along with confidence scores, encoded using five different reification approaches. 

Our goals for this dataset are twofold: First, we want to improve WDAqua-core0~\cite{diefenbach_wdaqua-core0:_2017} query answering system, providing it with more relations and lexicalizations, along with confidence scores that can help to give hints about how trustworthy is the answer. Second, given that it contains a big proportion of metadata statements, we want to use it as a testbed to compare how the different different metadata representations behave in current triplestores.

While currently we only publish the dumps of the datasets, we plan to provide SPARQL endpoint and full dereferenceable URLs. In addition, NELL is starting to be explored in languages different than English, such as Portuguese~\cite{hruschka_jr._coupling_2013,duarte_how_2014} and French~\cite{duarte_vers_2017}. Our intention is to convert those datasets to RDF as they become available to the public, since the system and knowledge base are exactly the same used in the English one.

\paragraph{Acknowledgements:}

This work is supported by funding from the EU H2020 research and innovation program under the Marie Sk\l{}odowska-Curie grant No 642795. We would like to thank Bryan Kisiel from NELL's CMU team for the technical support about NELL's components.

\bibliographystyle{mysplncsnat}
\bibliography{Zotero}

\end{document}